# Proportional Fair Resource Allocation on an Energy Harvesting Downlink - Part II: Algorithms

Neyre Tekbiyik, Tolga Girici, Elif Uysal-Biyikoglu, and Kemal Leblebicioglu

## Abstract

In this paper, the *proportionally fair* allocation of time slots in a frame, as well as power level to multiple receivers in an energy harvesting broadcast system, is considered. Energy harvest times in a frame are assumed to be known at the beginning of that frame. The goal is to solve an optimization problem designed to maximize a throughput-based utility function that provides proportional fairness among users. An optimal solution of the problem was obtained by using a Block Coordinate Descent (BCD) method in earlier work (presented in Part I of this study). However, finding the optimal allocation entails a computational complexity that increases sharply in terms of the number of users or slots. In this paper, certain structural characteristics of the optimal power-time allocation policy are derived. Building on those, two simple and computationally scalable heuristics, PTF and ProNTO are proposed. Simulation results suggest that PTF and ProNTO can closely track the performance of the BCD solution.

This work was supported by TUBITAK Grant 110E252.

N. Tekbiyik, E. Uysal-Biyikoglu, and Kemal Leblebicioglu are with the Department of Electrical and Electronics Engineering, Middle East Technical University, Ankara 06800 Turkey. e-mail: ntekbiyik,elif,lebleb@eee.metu.edu.tr

T. Girici is with the Department of Electrical and Electronics Engineering, TOBB Economics and Technology University, Ankara 06560 Turkey. e-mail: tgirici@etu.edu.tr

## I. INTRODUCTION

With increasing awareness of the potential harmful effects to the environment caused by "greenhouse gas" emissions and the depletion of non-renewable energy sources, there is a growing consensus on the need to develop more energy-efficient communication systems [1]. Whether the objective is to decrease the carbon footprint of wireless communications or to make nodes of a wireless network energy-wise self-sufficient, harvesting ambient energy is a promising approach for wireless communications. Ambient energy sources include sunlight, heat differentials, mechanical vibration, RF radiation, or any other physical source that can produce an electrical charge through a transducer (photovoltaic cell , piezoelectric element, etc.). Communication devices that can be powered by rechargeable batteries which store energy harvested through such means are already commercially available. However, harvested power is typically irregular and can at times fall short providing typical power consumption levels in wireless nodes. If it is desired for energy harvesting systems to match the performances their regular battery or grid-powered counterparts, the need to accomplish this by efficiently utilizing an unsteady power source opens up new challenges for the design of transmission as well as resource allocation schemes.

There has been a considerable amount of recent research effort on optimizing data transmission with an energy harvesting transmitter. A single-user communication system operating with an energy harvesting transmitter is considered in [2], where a packet scheduling scheme that minimizes the time by which all of the packets are delivered to the receiver is obtained. A multi-user extension of [2] has also been considered in [3], [4] and the same time minimization problem is solved for a two user broadcast channel. These approaches are exended in [5] and [6] to the case of a transmitter with a finite capacity battery. [7] extends [2] to propose the directional water-filling algorithm that finds the optimal energy management schemes for energy harvesting systems operating in fading channels, with finite capacity rechargeable batteries. In [8], the authors consider the problem of energy allocation over a finite horizon for point-to-point wireless communications and use dynamic programming and convex optimization techniques to obtain the throughput-optimal energy allocation.

The first part of our study on this problem [9] differs from the above-mentioned studies (and others cited in [9]) particularly in its aim to maximize the throughput in a *proportionally fair* way, taking into account the inherent differences of channel quality among users. In [9], we considered allocating among users the transmission power and the proportion of the time between energy harvests, to achieve a good balance between throughput and fairness on an energy harvesting downlink. Specifically, a proportional fairness based utility maximization problem in a time-sharing multi-user additive white Gaussian noise (AWGN) broadcast channel, where the transmitter is capable of energy harvesting is considered. The aim is to achieve the optimum *off-line* schedule, by assuming that the energy arrival profile at the transmitter is deterministic and known ahead of time in an off-line manner for a time window, called frame, i.e., the energy harvesting times and the corresponding harvested energy amounts are known at the beginning of each frame. The treatment in [9] considers the general case in which the interarrival times between consequtive harvests do not have to be equal. Here, we focus on the case where energy interarrival times are equal. Not all generality is lost, because harvest amounts are arbitrary and the absence of a harvest in a certain slot can be expressed with a harvest of amount zero for the respective slot. Periodic sampling of harvests is also consistent with practice as in many energy harvesting systems, transmitters have supercapacitors that can store the harvested energy and supply in every predetermined time window, allowing the case of periodic energy arrivals.

In this paper, we show that by using the periodic energy arrivals assumption, it is possible to analytically derive the characteristics of the optimal solution of the Problem proposed in [9]. In [9], we proved that the problem in hand is a biconvex problem and has multiple local optima. This allowed us to decompose the problem into two parts (power allocation, time allocation) and present a Block Coordinate Descent based optimization algorithm, BCD [9], that converges to a partial optimal solution. We believe that the partial optima found by BCD algorithm is very close to the local optima of the problem and thus, it achieves optimal or close to optimal utility [9]. Although BCD is guaranteed to converge to a partial optimal solution and thus the partial optimal utility, it is computationally expensive and when there are tens of users and energy arrivals, forming invertible hessian matrices (needed for the optimization of the power variables) may be computationally excessive. Hence in this paper, we first derive the characteristics of the optimal solution and then, build on those to develop simple heuristics, PTF and ProNTO that closely track the performance of the BCD solution.

We start by describing the system model in the next section. Next, we make the problem statement precise in Section III. Section IV discusses the structure and properties of the optimal solution. Depending on these properties, PTF and ProNTO heuristics are proposed in Section V. In Section VII, we present our numerical and simulation results. We conclude in Section VIII with an outline of future directions.

## II. SYSTEM MODEL

Consider a time-slotted system where each frame, of length $F_i$, is divided into $K$ slots. There is a single transmitter that transmits to $N$ users by time sharing. Channel conditions remain constant during $F_i$ ($g_n$, the gain of user $n$, is chosen to be constant throughout the frame). The transmitter is equipped with a rechargeable battery such that some energy, $E_{ti}$, is harvested from the environment at the beginning of each time slot $t$ of frame $i$. The length of the $t^{th}$ slot of frame $i$ may be represented as $T_{ti}$. However, as we are interested in a specific frame, we drop the frame indicator $i$ and define the harvested energy in slot $t$ as $E_t$, and, the length of slot $t$ as $T_t$. Note that, we use the same system model as in [9]. However, unlike [9], in this paper we assume periodic energy arrivals and hence equal slot lengths ($T_t = T$ for all $t = 1, \ldots, K$), as shown in Figure 1, to reveal the characteristics of the optimal solution of Problem 1.

Similar to the setting in [9], for a given frame, the transmitter chooses a power level $p_t$ and a time allocation vector $\tau_t = (\tau_{1t}, ..., \tau_{Nt})$, for each time slot $t$ of the frame, where $p_{nt} = p_t$ is the selected transmission power for user $n$ during slot $t$ and, $\tau_{nt}$ is the time allocated for transmission to user $n$ during slot $t$.

## III. PROBLEM STATEMENT

The total achievable number of bits sent to user $n$ within one frame (proportional to the *throughput* obtained by that user in the frame), is $\sum_{t=1}^{K} \tau_{nt} W \log_2 \left(1 + \frac{p_t g_n}{N_o W}\right)$ [9]. We aim to maximize a utility function, the log-sum throughput over the users $\sum_{n=1}^{N} log_2(R_n)$, which is known to achieve proportional fairness [10], in the presence of energy harvesting. We start with the problem of interest, Problem 1 defined in [9]. This is a constrained optimization problem that aims to maximize the utility function with respect to the time and energy constraints.

*Problem 1:*

$$\text{Maximize: } U(\overline{\tau}, \overline{p}) = \sum_{n=1}^{N} \log_2 \left( \sum_{t=1}^{K} \tau_{nt} W \log_2 \left(1 + \frac{g_n p_t}{N_o W}\right) \right)$$

$$\text{subject to: } \tau_{nt} \geq 0 \ , \ p_t \geq 0 \tag{1}$$

$$\sum_{n=1}^{N} \tau_{nt} = T \tag{2}$$

$$\sum_{t=1}^{K} \tau_{nt} \geq \epsilon \tag{3}$$

$$\sum_{i=1}^{t} p_i T_i \leq \sum_{i=1}^{t} E_i \tag{4}$$

where $t = 1, ..., K$ and $n = 1, ..., N$. $W$ is the bandwidth for a single link channel, and $N_o$ is the power spectral density of the background noise. Hence, $\frac{g_n p_t}{N_o W}$ represents the $SNR$ of user $n$ in slot $t$. Equations in (1) represent the nonnegativity constraints. The time-limit constraints, the set of equations in (2), ensure that the total time allocated to users does not exceed the slot length. The set of equations in (3), on the other hand, are technical constraints and ensure that every user gets some time ($\geq \epsilon$ where $\epsilon$ is an infinitely small number) during the frame. The set of equations in (4) are called energy causality constraints as these ensure no energy is transmitted before becoming available.

One might hope that this problem has a unique solution and no local optima except for one global optimum. Unfortunately, (1) is a nonlinear non-convex problem with potentially multiple local optima. Indeed, in [9], analysis of structural characteristics of the problem revealed that it can be formulated as a biconvex optimization problem, and that it has multiple optima. In the next section, we decompose Problem 1 into two parts (power allocation, time allocation) to investigate and derive the characteristics of these optima.

## IV. Structure and Properties of the Optimal Solution

In this section, we analyze the structure and properties of the hybrid power-time allocation policy. Remember that the utility function of Problem 1 is

$$U = \sum_{n=1}^{N} log_2(\sum_{t=1}^{K} \tau_{nt} R_{nt}) \tag{5}$$

where $R_{nt}$ represents the rate of link $n$ in $t^{th}$ slot:

$$R_{nt} = W log_2 \left(1 + L_n p_t\right) \quad where \quad L_n = \frac{g_n}{N_o W} \tag{6}$$

Let $\overline{R_n} = [R_{n1} \ R_{n2} \ \ldots \ R_{nK}]^T$ and $\overline{\tau_n} = [\tau_{n1} \ \tau_{n2} \ \ldots \ \tau_{nK}]^T$. Then, utility can be rewritten as

$$U = \sum_{n=1}^{N} log_2(\overline{\tau_n}^T \overline{R_n}) \tag{7}$$

$$= U_1 + U_2 + \ldots + U_N \tag{8}$$

where $U_n$, the utility of user $n$, is

$$U_n = log_2(\overline{\tau_n}^T \overline{R_n}) \quad (9)$$

In order to reveal characteristics related to the optimal solution that will help us develop computationally efficient and close-to-optimal heuristics, we decompose the problem into two parts (similarly as in [9]): power allocation and time allocation.

*A. Structure of an Optimal Power Allocation Policy*

In this section, we analyze the structure and properties of the optimal power allocation policy. In order to do this, we assume that the time allocation is determined, and try to characterize the structure of the optimal solution of the power allocation problem for this time allocation. Clearly, when the only variables are power variables, Problem 1 reduces to the following constrained optimization problem:

*Problem 2:*

$$\text{Maximize: } U(\overline{p}) = \sum_{n=1}^{N} U_n(\overline{p})$$

$$\text{subject to: } p_t \geq 0 \quad (10)$$

$$\sum_{i=1}^{t} p_i T_i \leq \sum_{i=1}^{t} E_i \quad (11)$$

where $t = 1, ..., K$ and, $U_n$ is a function of the power variables (as defined in Eq. (9)). In our previous work [9], we proved the strict convexity[1] of Problem 2. Similarly, the general problem, Problem 1, is shown to be a biconvex optimization problem that has many local minima [9]. As Problem 2 has a unique optimum, the optimal power allocation changes for every given time allocation. In Theorem 1, we claim that one of the optimum schedules of Problem 1 has a nondecreasing power schedule. Lemma 1 not only helps us to prove our claim but also reveals that Problem 1 has multiple optima. From the proof Lemma 1, the attentive reader can observe that any feasible permutation[2] of the optimal schedule $(\overline{\tau}^*, \overline{p}^*)$, described in Theorem 1, is also optimal.

---

[1] Maximizing $U(\overline{p})$ is equivalent to minimizing $-U(\overline{p})$ which is a convex objective function.

*Theorem 1:* When all slots have equal length ($T_j = T, \ for \ \forall j \in \{1,...,K\}$), there exists an optimal schedule $(\overline{\tau}^*, \overline{p}^*)$ such that $\overline{p}^*$ is nondecreasing, (e.g., $\overline{p}^* = (p_1,...,p_K)$ where $p_1 \leq p_2 \leq ... \leq p_K$).

*Proof:* The proof is provided in Appendix A, and rests on Lemma 1 below. ∎

We shall need the following definition of a permutation of a vector sorted in nondecreasing order of elements, for stating Lemma 1.

*Definition 1:* Given a vector $\overline{R_n} = [R_{n1} \ R_{n2} \ ... \ R_{nK}]^T$, we define $\overline{R_n}^\uparrow = [R_{n\pi(1)} \ R_{n\pi(2)} \ ... \ R_{n\pi(K)}]^T$ where $\overline{R_n}^\uparrow$ is a permutation (sorted in increasing order) of $\overline{R_n}$, such that

$$\overline{R_{n\pi(1)}} \leq ... \leq \overline{R_{n\pi(2)}} \leq ... \leq \overline{R_{n\pi(K)}} \quad (12)$$

*Lemma 1:* When all slots have equal length ($T_j = T, \ for \ \forall j \in \{1,...,K\}$), for any given schedule $(\overline{\tau}, \mathcal{P}_\mathcal{C})$, we can find such $\overline{\tau'_n}, \overline{R'_n}$ (where $\overline{R'_n} = \overline{R_n}^\uparrow$) that $(\overline{\tau'_n})^T \overline{R'_n} = \overline{\tau_n}^T \overline{R_n}$ for all $n = 1,...,N$; i.e., the utility, $U$, does not change. Hence, if $(\overline{\tau_n}^*, \overline{R_n}^*)$ is optimal, then $(\overline{\tau'_n}^*, \overline{R'_n}^*)$ is also optimal.

*Proof:* The proof is provided in Appendix B. ∎

## B. Structure of an Optimal Time Allocation Policy

In this section, we assume that the power allocation through the slots is determined. Then, given that the power variables are known constants, we determine the structure and properties

---

[2] A feasible permutation is any permutation of a given schedule that does not violate the constraints described in Eqns. (1)-(4).

of the optimal time allocation policy. Since, the only variables are time variables, Problem 1 reduces to Problem 3:

---

*Problem 3:*

$$\text{Maximize: } U(\overline{\tau}) = \sum_{n=1}^{N} U_n(\overline{\tau})$$

$$\text{subject to: } \tau_{nt} \geq 0 \tag{13}$$

$$\sum_{n=1}^{N} \tau_{nt} = T \tag{14}$$

$$\sum_{t=1}^{K} \tau_{nt} \geq \epsilon \tag{15}$$

---

where $t = 1, ..., K$, $n = 1, ..., N$ and, $U_n$ is a function of the time variables (as defined in Eq. (9)). In [9], Problem 3 is shown to be convex. Thus, the analysis can rely on KKT (Karush-Kuhn-Tucker) optimality conditions, which must be satisfied by the global optimum. We start by forming the Lagrangian function as follows:

$$L(\overline{\tau}, \overline{\lambda}, \overline{\mu}) = -U(\overline{\tau}) + \sum_{j=1}^{K}\sum_{i=1}^{N} \mu_{(N(j-1)+i)}\tau_{ij} + \sum_{j=NK+1}^{NK+N} \mu_j(\epsilon - \sum_{t=1}^{K} \tau_{(j-NK)t}) + \sum_{i=1}^{K} \lambda_i(\sum_{n=1}^{N} \tau_{ni} - T_i) \tag{16}$$

where $\mu$'s are the Lagrange multipliers, and, the total number of constraints[3] is $N(K+1)+K$. After defining the Lagrangian as in Eq. (16), one can construct the KKT conditions for the optimal solution. Due to space limitations, we do not list the conditions here but refer the interested reader to the associated technical report [11] for the details. Please note that the optimal time allocation should jointly satisfy the set of equations that arise from KKT conditions. Clearly, as the number of users, $N$, and, the number of slots, $K$, increase, the number of equations increases dramatically making it cumbersome to write analytical solutions. Therefore, for the sake of conciseness, we continue the analysis with the special case of two users and two slots which allows us to construct the characteristics of the optimal time allocation policy.

---

[3]There are $K$ equality constraints and $NK + N$ inequality constraints.

Consider two consequtive slots with different power levels. Let us call the one with the least power *the weak slot*, and the one with the highest power *the strong slot*. When the slots have equal length ($T_1 = T_2 = T$), the optimal policy has the properties described in Lemma 2.

*Lemma 2:* In an optimal schedule, time allocation over the two slots (of equal length) has the following properties:

1) The weak slot is assigned to only one of the users. The strong slot, however, is shared between users. When both power levels are equal; if one slot is assigned to user 1 (user 2), the other slot is assigned to user 2 (user 1).

2) To whom the the weak slot will be assigned depends on two criteria: first, $\Gamma_n = \frac{R_{n2}}{R_{n1}}$, which is the ratio of user $n$'s rate in the second slot to that in the first, and second, whether the strong slot is before or after the weak slot. When the weak slot preceeds the strong slot, it is assigned to the user with the smaller $\Gamma$. Otherwise (implying the decrease in power level), it is assigned to the user with the higher $\Gamma$.

3) In a strong slot, the user that did not (or will not) receive any data in the weak slot is favored, i.e., more than half of the slot is assigned to that user. In order to preserve fairness, this favoring operation is done by considering $\Gamma_1$ and $\Gamma_2$.

*Proof:* The proof is provided in Appendix C. ∎

## V. PTF Heuristic

In this section, we develop a heuristic algorithm, Power-Time-Fair (PTF), based on the characteristics (discovered in the previous section) of an optimal power/time allocation schedule. The PTF algorithm operates as follows:

1) **For Power Allocation:** Assign nondecreasing powers through the slots by using the energy

harvest statistics, as follows:

   a) From a slot, say $i$, to the next one $i+1$: If harvested energy decreases, defer a $\Delta$ amount of energy from slot $i$ to slot $i+1$ to equalize the power levels. Do this until all powers are nondecreasing, and, form a virtual nondecreasing harvest order.

   b) By using the virtual harvest order, assign nondecreasing powers through the slots, i.e., in each slot, spend what you virtually harvested at the beginning of that slot.

2) **For Time Allocation:** For the power allocation found in 1), let, $B_{nt} = R_{nt}T$ be the number of bits that would be sent by user $n$ if the whole slot (of length $T$) was allocated to that user. Assign the first slot to the user who has the maximum rate, $R_{nt}$, in that slot. For the other slots, apply the following: At the beginning of each slot, $t \in \{2, \ldots, K\}$, determine the user with the maximum $\beta$ where,

$$\beta_n = \frac{B_{nt}}{\sum_{i=1}^{t} B_{ni}}$$

and, assign the whole slot to that user. If multiple users share the same $\beta$, then, allocate the slot to the user with the best channel.

Simulation results show that the performance of the PTF algorithm is close to the performance of the BCD algorithm.

## VI. PRONTO HEURISTIC

In this section, we develop a fast and simple heuristic, ProNTO (Powers Nondecreasing - Time Ordered), based on the optimal power allocation related characteristics discovered in Section IV-A and the simulation results obtained by running BCD algorithm for periodic energy arrivals. The ProNTO algorithm operates as follows:

1) **For Power Allocation:** Assign nondecreasing powers through the slots by using the energy harvest statistics, as done in part (1) of PTF algorithm.

2) **For Time Allocation:** Order the users, $u_1, \ldots, u_N$, according to their channel quality and form a user priority vector, $\overline{u^\downarrow} = [u_1^\downarrow, \ldots, u_N^\downarrow]$ where $u_1^\downarrow$ represents the user with the best channel. As $K > N$, Allocate every user $\frac{K-mod(K,N)}{N}$ slots as follows: The first $\frac{K-mod(K,N)}{N}$ slots are allocated to $u_1^\downarrow$, the next $\frac{K-mod(K,N)}{N}$ slots are allocated to $u_2^\downarrow$, etc. Add the remaining $mod(K,N)$ slots to the most powerful $mod(K,N)$ users' slots. For example; Let $K = 12$ and $N = 5$, and the path losses of the users to be 13 dB, 17 dB, 10 dB, 12 dB, 20 dB respectively. Then, the first 3 slots are allocated to user 3, the next 3 slots are allocated to user 4, the following 2 slots are allocated to user 1, $9^{th}$ and $10^{th}$ slots are allocated to user 2, and the last 2 slots are allocated to user 5.

Thus PTF and ProNTO differ only in time allocation part. The time allocation method used in ProNTO is proposed according to the following observation: when a partial optimal solution obtained by BCD algorithm is modified as described in Lemma 1 and its proof, to form the nondecreasing optimal schedule, the time allocation becomes ordered, e.g., as shown in Table III. As time allocation method used in ProNTO is simpler than the one used in PTF, ProNTO can operate faster. Simulation results show that the performance of ProNTO is close to the performance of the BCD algorithm.

## VII. NUMERICAL AND SIMULATION RESULTS

In this section, we present the numerical and simulation results related to PTF and ProNTO heuristics. Throughout our simulations, we use the folowing setup: $W = 1kHz$, $N_o = 10^{-6}W/Hz$. We assume that some amount of energy ($\epsilon < E < \infty$ where $\epsilon$ is an infinitely small value) is harvested every 10 seconds ($T = 10$), within a frame (period of known harvests). Note that, throughout this section, the units used for frame length, energy, and power are; seconds, Joules, and Watts respectively. Throughout our simulations, we use four different frame lengths; 20, 80, 100, 120. For the frame of 20 secs, we use three different energy harvest models; $[0.5, 50]$, $[50, 0.5]$, $[6020]$. We define different cases for the remaining three frame lengths;

*Regular*, *Bursty*, and, *Very Bursty*. In *Regular*, the harvest amounts are close to each other and form a regular pattern; $E_R = [73, 65, 9, 19, 40, 37, 22, 84, 39, 67, 81, 100]$. In *Bursty*, there are short term sudden decreases and increases in harvest amounts, causing a bursty pattern; $E_B = [20, 100, 1, 1, 1, 70, 100, 1, 10, 40]$. Finally, *Very Bursty* represents an extreme case where the transmitter stays energy-hungry for a long time; $E_V = [90, 2, 0.5, 0.1, 0.3, 0.7, 40, 60]$.

We start by the simplest case of two users and two slots ($N = 2$, $K = 2$, frame of 20 secs) to compare the results obtained by BCD algorithm [9], with the optimal ones presented in Table I. Our objective in doing such a comparison is to prove the accuracy of both theoretical and simulation results. We refer the interested reader to Appendix C for the details of the optimality table, and provide the comparison in Table II. Note that as in [9], the starting point of the algorithm is the Spend What You Get (SG) policy (proposed by Gorlatova et. al. [12]) combined with TDMA time allocation (SG+TDMA). The first column of Table II shows the amount of the harvests ($E_1$, $E_2$). The second column represents the mean path loss (in dB) of the two users. As observed from the table, for a given power allocation, the results found by BCD algorithm and the optimal ones (obtained by KKT optimality conditions) are almost the same, verifying the consistency and optimality of the algorithm.

The attentive reader can observe from Table II that, when harvests decrease from one slot to another, the optimal powers tend to be nondecreasing. Hence in that case, the algorithm seems to be converged to the nondecreasing optimal discussed in Theorem 1. Note that, this nondecreasing optimal could also be obtained by using the modification method explained in Lemma 1. By using that method, we modify the results obtained by BCD algorithm to reveal the optimal (nondecreasing) power and time allocation policies for increasing number of users. For our analysis, we use three different path loss patterns, called, *Low*, *Moderate*, *High* respectively. In *Low*, the strongest user in the system has 13 dB path loss, and, every new user that joins the system deviates by 3 dB from the previous one (has 3 dB more path loss than the preceding user). In *Moderate*, the strongest user has 19 dB path loss, and, every new user deviates by 3 dB. Finally, in *High*, the strongest user has 25 dB path loss, and, every new user deviates 3 dB. Due to space limitations, we present only the *Bursty-Moderate* case's results in Table III. As illustrated, when the number of users increase, BCD algorithm tends to assign increasing powers rather than nondecreasing. One can also see from the table that, no matter how many users exist in the system, ordering powers in nondecreasing order, causes the time allocation to be ordered too. By ordered, we mean that the first slot(s) are allocated to the user with the best channel, the next slot(s) are allocated to the user with the second best channel, etc. , and the last slot(s) are allocated to the user with the worst channel. This observation constitutes the main motivation for the ProNTO heuristic.

We next use the above-mentioned energy harvesting cases (*Regular*, *Bursty*, *Very Bursty*) to compare the PTF and ProNTO heuristics' performances to that of BCD's. We start by testing the utility and throughput improvement (over SG+TDMA) performances of the heuristics for increasing path losses. For this, we set the number of users to two, i.e., $N = 2$. The results are presented in Figure 2 and Figure 3, respectively. In both figures, the Mean Path Loss, is computed as $\widetilde{L} = \frac{1}{N} \sum_{i=1}^{N} L_i$ where $L_i$ represents the path loss of user $i$. Hence, the three mean path losses seen in the figures represent the *Low*, *Moderate* and *High* cases. One can observe from Figure 2 that, the utility improvements of all algorithms tend to increase (or at least stay constant) when path loss increases, and the utility improvement performances of the proposed heuristics are very close to that of BCD's. For the chosen cases, ProNTO outperforms PTF. This is more obvious for the *Very Bursty* case. The corresponding throughput improvements are shown in Figure 3. As illustrated, for the case of $N = 2$, even with $\approx 5\%$ of utility improvement, a $\approx 65\%$ of improvement in total throughput is possible. Note that, in all cases, the performances are very close to each other.

In order to determine the effect of number of users to the performances of our proposed heuristics, we next perform a series of simulations by considering all energy harvesting cases (*Regular*, *Bursty*, *Very Bursty*) and different number of users. By taking average over all energy harvesting cases, we present the average utility improvement results in Figure 4, for the *Moderate* case. As illustrated in the figure, when the number of users increase, the average utility improvements of all schemes also increase. Note that, both heuristics closely track the BCD algorithm. When there are few users in the system, PTF and ProNTO are competitive. However, when there are more users, ProNTO seems to outperform PTF in terms of average utility improvement. At all instances, ProNTO is within the 1% neighbourhood of the BCD algorithm.

Although we aim at proportional fairness in this work, it may be interesting to analyse max-min fairnesses of the proposed algorithms, PTF and ProNTO. Jain's index is a well-known measure of fairness [13], [14]. The index $FI$ takes the value of 1 when there is a complete fair allocation.

$$FI = \frac{(\sum_{i=1}^{N} x_i)^2}{N \cdot \sum_{i=1}^{N} x_i^2} \qquad (17)$$

For computing $FI$, we use the no. of bits transmitted to the users, $x_i = 2^{U_i}$ for $i = 1, \ldots, N$, where $U_i$ is as defined in Eq. (9). From Table IV, it is clear that SG+TDMA is the worst choice in terms of fairness. Although low path losses embrace lower utility improvement, they mainly allow both PTF and ProNTO to be very efficient in terms of fairness. However, as observed from the table, when all three cases are considered, PTF seems to be more fair than ProNTO is. Hence, ProNTO seems to trade of fairness for utility improvement. It can also be inferred from Figure 4 and Table IV that, when ProNTO outperforms PTF in terms of utility improvement, the difference between two heuristics is not high. However, this is not the case for fairness,.i.e., when PTF outperforms ProNTO, the difference can be considered as high. Hence, although ProNTO seems more promising in terms of utility improvement, depending on system requirements, one can still choose PTF over ProNTO for more fairness.

## VIII. CONCLUSION

This paper presented the second part of a study whose first part was reported in [9]. Building on the problem formulation and the optimal solution method in [9], the optimal resource allocation policy was further studied and certain structural characteristics of the optimal solution were established. In particular, the existence of an optimal nondecreasing power schedule and, an ordered time allocation schedule were proved. This allowed us to propose two alternative efficient and scalable heuristics, PTF and ProNTO. The computational ease of these algorithms were observed in numerical examples, while the policies they result in coincide with the structural properties we have shown the optimal to have. Simulation results indicate that, despite their simplistic design, PTF and ProNTO heuristics can closely track the performance of the optimal BCD algorithm. In our examples, which were computed for small or moderate problem sizes, both PTF and ProNTO took one or two orders of magnitude smaller time to converge than BCD, which has to compute a Hessian. Typically, ProNTO outperforms PTF in terms of utility improvement, whereas the latter is fairer. The utility improvement difference between BCD and ProNTO is shown to be less than 1% at all instances.

An interesting future direction could be the development of an online algorithm that will bypass the need for offline knowledge about the energy harvesting statistics. This algorithm may use energy harvesting prediction algorithms to predict the energy that will arrive in the future, or estimate it on the fly during network operation.

## APPENDIX A

### PROOF OF THEOREM 1

The proof is done by contradiction. For any given time allocation $\overline{\tau}$, consider a given power sequence, $\mathcal{P}_\mathcal{C} = (p_1, ..., p_{d-1}, p_d, ..., p_K)$, in which the power level decreases at some time, say $d > 1$. In such a case, we can defer some energy, $0 < \Delta \leq p_{d-1}T_{d-1}$, from the $(d-1)^{th}$ slot to the $d^{th}$ slot forming a modified schedule, $\mathcal{P}'_\mathcal{C} = (p_1, ..., p'_{d-1}, p'_d, ..., p_K)$, that will not violate the energy causality conditions (as shown in Fig. 5). Clearly, we can continue this deferral operation until $p'_{d-1} < p'_d$ and still not violate the energy causality conditions. Applying the same method for every possible decrease leads us to a nondecreasing schedule, $\mathcal{P}^{\uparrow}_{\mathcal{C}} = (p'_1, ..., p'_{d-1}, p'_d, ..., p'_K)$, where $p'_1 \leq p'_2 \leq ... \leq p'_K$.

From Lemma 1, $U(\overline{\tau}, \mathcal{P}_{\mathcal{C}}) = U(\overline{\tau}^{\mathcal{P}^{\uparrow}_{\mathcal{C}}}, \mathcal{P}^{\uparrow}_{\mathcal{C}})$. Thus, for time allocation $\overline{\tau}^* = \overline{\tau}^{\mathcal{P}^{\uparrow}_{\mathcal{C}}}$, $\mathcal{P}^{\uparrow}_{\mathcal{C}}$ is optimal. This completes the proof.

APPENDIX B

PROOF OF LEMMA 1

Let, $\overline{R'_n} = \overline{R_n}^{\uparrow}$ where $\overline{R_n}^{\uparrow}$ is as defined in Definition 1. Note that Eq. (12) forces

$$log_2(1 + L_n p'_1) \leq \ldots \leq log_2(1 + L_n p'_l) \leq \ldots \leq log_2(1 + L_n p'_K) \tag{18}$$

$$1 + L_n p'_1 \leq \ldots \leq 1 + L_n p'_l \leq \ldots \leq 1 + L_n p'_K \tag{19}$$

$$p'_1 \leq \ldots \leq p'_l \leq \ldots \leq p'_K \tag{20}$$

Hence, sorting $\overline{R_n}$ in increasing order, forces nondecreasing powers (ordered schedule $\mathcal{P}^{\uparrow}_{\mathcal{C}}$ mentioned previously), which indeed forces all other $\overline{R_i}$ (where $i \in \{1, \ldots, i-1, i+1, \ldots, N\}$) to be sorted in increasing order, to form $\overline{R'_i}$. Now, we have new rates, $\overline{R'_i}$ for all users $i = 1, \ldots, N$. Remember that the utility of a user is defined as in Eq. (9). Thus, changing the order of $\overline{R_i}$ vector does not change the value of $U_i$ if the order of $\overline{\tau_i}$ is also changed so that the previous element pairs are matched again. Let us explain this, with an example. Let $R_{i2} < R_{i1}$, $R_{NK} < R_{i2}$, and, $R_{i1} \leq R_{i3} \leq \ldots \leq R_{i(K-1)}$. Then, $\overline{\tau'_i}$, and, $\overline{R'_i}$ vectors are defined as $\overline{R'_i} = [R_{iK} \ R_{i2} \ R_{i1} \ R_{i3} \ \ldots \ R_{i(K-1)}]^T$ and $\overline{\tau'_i} = [\tau_{iK} \ \tau_{i2} \ \tau_{i1} \ \tau_{i3} \ \ldots \ \tau_{i(K-1)}]^T$. Hence, it is straight forward to write that

$$\overline{\tau_i'}^T \overline{R_i'} = \tau_{iK} R_{iK} + \tau_{i2} R_{i2} + \tau_{i1} R_{i1} + \ldots + \tau_{i(K-1)} R_{i(K-1)}$$

$$= \tau_{i1} R_{i1} + \tau_{i2} R_{i2} + \ldots + \tau_{i(K-1)} R_{i(K-1)} + \tau_{iK} R_{iK}$$

$$= \overline{\tau_i}^T \overline{R_i} \tag{21}$$

where $\overline{\tau_i}$ and $\overline{R_i}$ are as defined in Eq. (7). As it can be observed, $U_i = U_i'$ as long as $\overline{R_i'} = \overline{R_i}^\uparrow$ and $\overline{\tau_i'} = (\overline{\tau_i})^{\overline{R_i}^\uparrow}$. Here, $\overline{\tau_i}^{\overline{R_i}^\uparrow}$ indicates the $\overline{\tau_i}$ vector ordered according to $\overline{R_i}^\uparrow$. Under these circumstances, $U_i = U_i'$ for all $i = 1, \ldots, N$, and, the overall utility does not change, $U = U'$. This completes the proof.

## APPENDIX C

### PROOF OF LEMMA 2

For the proof of Lemma 2, we use the KKT optimality conditions. Let, $A_n = \tau_{n1}^* R_{n1} + \tau_{n2}^* R_{n2}$. Then, for the special case, ($N = 2$, $K = 2$), the set of KKT conditions described in [11] reduces to Eqns. (22a)-(22g).

$$\frac{\partial L}{\partial \tau_{nt}} = \frac{1}{ln2} \frac{R_{nt}}{A_n} + \mu_{2(t-1)+n}^* + \mu_{n+4}^* - \lambda_t^* = 0 \tag{22a}$$

$$\mu_i^* \geq 0 \tag{22b}$$

$$\tau_{nt}^* \geq 0 \tag{22c}$$

$$\tau_{n1}^* + \tau_{n2}^* \geq \epsilon \tag{22d}$$

$$\tau_{1t}^* + \tau_{2t}^* = T \tag{22e}$$

$$\mu_{2(t-1)+n}^* \tau_{nt}^* = 0 \tag{22f}$$

$$\mu_{4+n}^*(\tau_{n1}^* + \tau_{n2}^* - \epsilon) = 0 \tag{22g}$$

for $i = 1, \ldots, 6$, $n = 1, 2$ and $t = 1, 2$. Combining the set of equations described above leads us to the following optimality conditions for the time allocation:

$$\mu_{2t-1}^* \tau_{1t}^* = 0 \tag{23a}$$

$$\left(\frac{R_{1t}}{A_1 ln2} - \frac{R_{2t}}{A_2 ln2} + \mu_{2t-1}^*\right)(T - \tau_{1t}^*) = 0 \tag{23b}$$

Solving the set of equations in Eq. (23), one can obtain the desired relation between power allocation and time allocation, as illustrated in Table I. Due to the convex nature of the problem, the solutions presented in Table I represent the global optima, when the rate improvements of the users, $\Gamma_n$, are equal. By inspecting Table I, one can observe the properties mentioned in Lemma 2. All cases are summarized in Table I, which completes the proof.

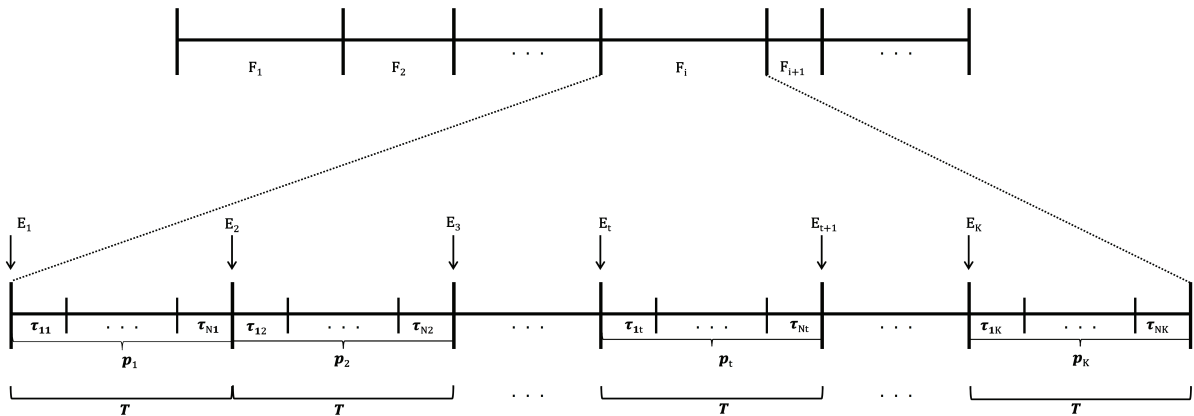

Fig. 1. Problem illustration: There are $K$ energy arivals in a frame, and, the time between consecutive arrivals are allocated to $N$ users.

TABLE I
OVERALL OPTIMALITY CONDITIONS FOR THE SPECIAL CASE OF TWO USERS AND TWO SLOTS ($T_1 = T_2$): CATEGORIZED ACCORDING TO THE RELATION BETWEEN THE POWERS ALLOCATED IN THE FIRST AND SECOND SLOTS. FOR A GIVEN POWER ALLOCATION, THE OPTIMAL TIME ALLOCATION DIFFERS ACCORDING TO THE RELATION BETWEEN THE RATE IMPROVEMENTS OF THE USERS.

| Power Relation (Slot 1 vs. Slot 2) | Users' Rate Improvement Relation (User 1 vs. User 2) | Slot 1 User 1 $\tau_{11}$ | Slot 1 User 2 $\tau_{21}$ | Slot 2 User 1 $\tau_{12}$ | Slot 2 User 2 $\tau_{22}$ | Utility |
|---|---|---|---|---|---|---|
| $p_1 < p_2$ | $\Gamma_1 < \Gamma_2$ | $T$ | $0$ | $\frac{T}{2}\left(1 - \frac{1}{\Gamma_1}\right)$ | $\frac{T}{2}\left(1 + \frac{1}{\Gamma_1}\right)$ | $\log_2\left(\frac{R_{22}}{R_{12}}(R_{11} + R_{12})^2\right) + 2\log_2\left(\frac{T}{2}\right)$ |
| | $\Gamma_1 = \Gamma_2$ | $T$ | $0$ | $\frac{T}{2}\left(1 - \frac{1}{\Gamma_1}\right)$ | $\frac{T}{2}\left(1 + \frac{1}{\Gamma_1}\right)$ | $\log_2\left((R_{11} + R_{12})(R_{21} + R_{22})\right) + 2\log_2\left(\frac{T}{2}\right)$ |
| | $\Gamma_1 = \Gamma_2$ | $0$ | $T$ | $\frac{T}{2}\left(1 + \frac{1}{\Gamma_2}\right)$ | $\frac{T}{2}\left(1 - \frac{1}{\Gamma_2}\right)$ | $\log_2\left((R_{11} + R_{12})(R_{21} + R_{22})\right) + 2\log_2\left(\frac{T}{2}\right)$ |
| | $\Gamma_1 > \Gamma_2$ | $0$ | $T$ | $\frac{T}{2}\left(1 + \frac{1}{\Gamma_2}\right)$ | $\frac{T}{2}\left(1 - \frac{1}{\Gamma_2}\right)$ | $\log_2\left(\frac{R_{12}}{R_{22}}(R_{21} + R_{22})^2\right) + 2\log_2\left(\frac{T}{2}\right)$ |
| $p_1 = p_2$ | $\Gamma_1 < \Gamma_2$ | $T$ | $0$ | $0$ | $T$ | $\log_2(R_{11}R_{22}) + 2\log_2(T)$ |
| | $\Gamma_1 = \Gamma_2$ | $T$ | $0$ | $0$ | $T$ | $\log_2(R_{11}R_{22}) + 2\log_2(T)$ |
| | $\Gamma_1 = \Gamma_2$ | $0$ | $T$ | $T$ | $0$ | $\log_2(R_{11}R_{22}) + 2\log_2(T)$ |
| | $\Gamma_1 > \Gamma_2$ | $0$ | $T$ | $T$ | $0$ | $\log_2(R_{12}R_{21}) + 2\log_2(T)$ |
| $p_1 > p_2$ | $\Gamma_1 < \Gamma_2$ | $\frac{T}{2}(1 + \Gamma_2)$ | $\frac{T}{2}(1 - \Gamma_2)$ | $0$ | $T$ | $\log_2\left(\frac{R_{11}}{R_{21}}(R_{21} + R_{22})^2\right) + 2\log_2\left(\frac{T}{2}\right)$ |
| | $\Gamma_1 = \Gamma_2$ | $\frac{T}{2}(1 - \Gamma_2)$ | $\frac{T}{2}(1 + \Gamma_2)$ | $T$ | $0$ | $\log_2\left((R_{11} + R_{12})(R_{21} + R_{22})\right) + 2\log_2\left(\frac{T}{2}\right)$ |
| | $\Gamma_1 = \Gamma_2$ | $\frac{T}{2}(1 + \Gamma_1)$ | $\frac{T}{2}(1 - \Gamma_1)$ | $0$ | $T$ | $\log_2\left((R_{11} + R_{12})(R_{21} + R_{22})\right) + 2\log_2\left(\frac{T}{2}\right)$ |
| | $\Gamma_1 > \Gamma_2$ | $\frac{T}{2}(1 - \Gamma_1)$ | $\frac{T}{2}(1 + \Gamma_1)$ | $T$ | $0$ | $\log_2\left(\frac{R_{21}}{R_{11}}(R_{11} + R_{12})^2\right) + 2\log_2\left(\frac{T}{2}\right)$ |

TABLE II
BCD VS. OPTIMAL RESULTS FOR THE SPECIAL CASE OF TWO USERS AND TWO SLOTS

| Harvests | Mean Path loss | Power Allocation by BCD | Time Allocation by BCD | Optimal Time Allocation | Utility by BCD | Optimal Utility |
|---|---|---|---|---|---|---|
| [0.5 50] | 20.5 | [0.0500  5.0000] | $\begin{bmatrix} 10 & 4.4129 \\ 0 & 5.5871 \end{bmatrix}$ | $\begin{bmatrix} 10 & 4.4129 \\ 0 & 5.5871 \end{bmatrix}$ | 29.8094 | 29.8094 |
| [0.5 50] | 26.5 | [0.0500  5.0000] | $\begin{bmatrix} 10 & 4.7399 \\ 0 & 5.2601 \end{bmatrix}$ | $\begin{bmatrix} 10 & 4.7399 \\ 0 & 5.2601 \end{bmatrix}$ | 28.4062 | 28.4062 |
| [0.5 50] | 32.5 | [0.0500  5.0000] | $\begin{bmatrix} 10 & 4.8786 \\ 0 & 5.1214 \end{bmatrix}$ | $\begin{bmatrix} 10 & 4.8786 \\ 0 & 5.1214 \end{bmatrix}$ | 26.5152 | 26.5152 |
| [50 0.5] | 20.5 | [2.2993  2.7507] | $\begin{bmatrix} 10 & 0.2428 \\ 0 & 9.7572 \end{bmatrix}$ | $\begin{bmatrix} 10 & 0.2431 \\ 0 & 9.7569 \end{bmatrix}$ | 30.9401 | 30.9401 |
| [50 0.5] | 26.5 | [2.2466  2.8034] | $\begin{bmatrix} 10 & 0.4291 \\ 0 & 9.5709 \end{bmatrix}$ | $\begin{bmatrix} 10 & 0.4295 \\ 0 & 9.5705 \end{bmatrix}$ | 29.4618 | 29.4618 |
| [50 0.5] | 32.5 | [2.2110  2.8390] | $\begin{bmatrix} 10 & 0.7024 \\ 0 & 9.2976 \end{bmatrix}$ | $\begin{bmatrix} 10 & 0.7047 \\ 0 & 9.2953 \end{bmatrix}$ | 27.2580 | 27.2580 |
| [60 20] | 2.5 | [3.8238  4.1762] | $\begin{bmatrix} 10 & 0.0538 \\ 0 & 9.9462 \end{bmatrix}$ | $\begin{bmatrix} 10 & 0.0544 \\ 0 & 9.9456 \end{bmatrix}$ | 33.5272 | 33.5272 |
| [60 20] | 8.5 | [3.7879  4.2121] | $\begin{bmatrix} 10 & 0.0786 \\ 0 & 9.9214 \end{bmatrix}$ | $\begin{bmatrix} 10 & 0.0787 \\ 0 & 9.9213 \end{bmatrix}$ | 32.9577 | 32.9577 |
| [60 20] | 14.5 | [3.7379  4.2621] | $\begin{bmatrix} 10 & 0.1216 \\ 0 & 9.8784 \end{bmatrix}$ | $\begin{bmatrix} 10 & 0.1216 \\ 0 & 9.8784 \end{bmatrix}$ | 32.2496 | 32.2496 |

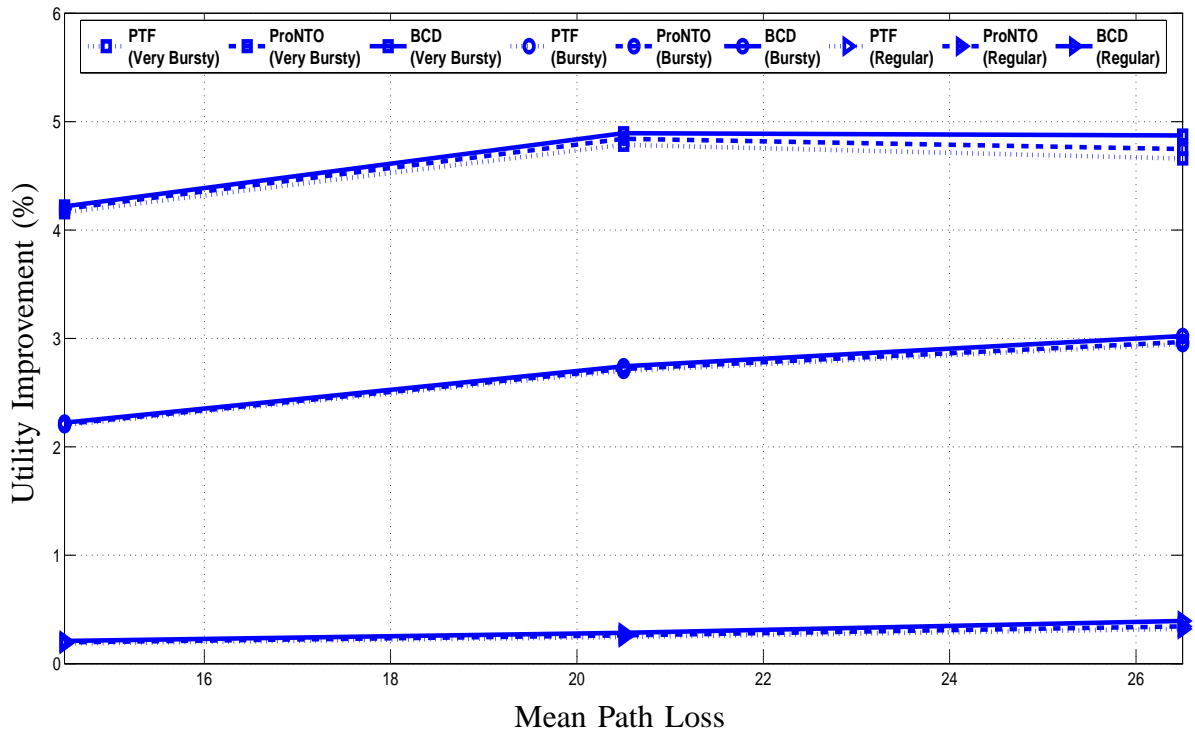

Fig. 2. Utility Improvement (BCD, PTF, ProNTO) vs. Mean Path Loss for $N = 2$: The effect of mean path loss on utility improvement for the three energy harvesting cases; *Regular*, *Bursty*, *Very Bursty*.

TABLE III
OPTIMAL TIME AND POWER ALLOCATION POLICIES VS. NUMBER OF USERS: FOUND BY BCD ALGORITHM AND MODIFIED ACCORDING TO LEMMA 1

| No. of Users | T.A. / P.A. | Users/Slots | Slot 1 | Slot 2 | Slot 3 | Slot 4 | Slot 5 | Slot 6 | Slot 7 | Slot 8 | Slot 9 | Slot 10 |
|---|---|---|---|---|---|---|---|---|---|---|---|---|
| 2 | T.A. | 1 | 10 | 10 | 10 | 10 | 10 | 3.6666 | 0 | 0 | 0 | 0 |
|   |      | 2 | 0  | 0  | 0  | 0  | 0  | 6.3334 | 10 | 10 | 10 | 10 |
|   | P.A. |   | 2  | 2.5750 | 2.5750 | 2.5750 | 2.5750 | 4.2117 | 4.4720 | 4.4720 | 4.4720 | 4.4720 |
| 3 | T.A. | 1 | 10 | 10 | 10 | 7.1883 | 0 | 0 | 0 | 0.0128 | 0 | 0 |
|   |      | 2 | 0  | 0  | 0  | 2.8117 | 10 | 10 | 10 | 9.9872 | 0 | 0 |
|   |      | 3 | 0  | 0  | 0  | 0 | 0 | 0 | 0 | 0 | 10 | 10 |
|   | P.A. |   | 2  | 2.4151 | 2.4151 | 2.5531 | 2.9166 | 3.9160 | 4 | 4.7272 | 4.7283 | 4.7283 |
| 4 | T.A. | 1 | 10 | 10 | 8.5777 | 0 | 0 | 0 | 0 | 0 | 0 | 0 |
|   |      | 2 | 0  | 0  | 1.4223 | 10 | 10 | 4.7598 | 0 | 0 | 0 | 0 |
|   |      | 3 | 0  | 0  | 0 | 0 | 0 | 5.2401 | 10 | 8.3789 | 0 | 0 |
|   |      | 4 | 0  | 0  | 0 | 0 | 0 | 0 | 0 | 1.6211 | 10 | 10 |
|   | P.A. |   | 2  | 2.3132 | 2.3810 | 2.8028 | 2.8028 | 3.6343 | 4.0501 | 4.2070 | 5.0742 | 5.0742 |
| 5 | T.A. | 1 | 10 | 10 | 3.3321 | 0 | 0 | 0 | 0 | 0 | 0 | 0 |
|   |      | 2 | 0  | 0  | 6.6678 | 10 | 5.1071 | 0 | 0 | 0 | 0 | 0 |
|   |      | 3 | 0  | 0  | 0 | 0 | 4.8928 | 10 | 5.2864 | 0 | 0 | 0 |
|   |      | 4 | 0  | 0  | 0 | 0 | 0 | 0 | 4.7135 | 10 | 3.4256 | 0 |
|   |      | 5 | 0  | 0  | 0 | 0 | 0 | 0 | 0 | 0 | 6.5743 | 10 |
|   | P.A. |   | 2  | 2.1351 | 2.4281 | 2.5792 | 3.1523 | 3.1574 | 3.8909 | 4.3727 | 5.1247 | 5.5592 |
| 6 | T.A. | 1 | 10 | 10 | 0.5333 | 0 | 0 | 0 | 0 | 0 | 0 | 0 |
|   |      | 2 | 0  | 0  | 9.4666 | 10 | 0 | 0 | 0 | 0 | 0 | 0 |
|   |      | 3 | 0  | 0  | 0 | 0 | 10 | 7.3867 | 0 | 0 | 0 | 0 |
|   |      | 4 | 0  | 0  | 0 | 0 | 0 | 2.6132 | 10 | 3.3768 | 0 | 0 |
|   |      | 5 | 0  | 0  | 0 | 0 | 0 | 0 | 0 | 6.6231 | 7.7727 | 0 |
|   |      | 6 | 0  | 0  | 0 | 0 | 0 | 0 | 0 | 0 | 2.2272 | 10 |
|   | P.A. |   | 1.8639 | 1.8639 | 2.2338 | 2.2553 | 3.0596 | 3.2585 | 3.8704 | 4.0829 | 5.3135 | 6.5977 |
| 7 | T.A. | 1 | 10 | 8.9882 | 0 | 0 | 0 | 0 | 0 | 0 | 0 | 0 |
|   |      | 2 | 0  | 1.0117 | 10 | 6.4262 | 0 | 0 | 0 | 0 | 0 | 0 |
|   |      | 3 | 0  | 0 | 0 | 3.5737 | 10 | 2.4354 | 0 | 0 | 0 | 0 |
|   |      | 4 | 0  | 0 | 0 | 0 | 0 | 7.5645 | 6.7008 | 0 | 0 | 0 |
|   |      | 5 | 0  | 0 | 0 | 0 | 0 | 0 | 3.2991 | 10 | 0 | 0 |
|   |      | 6 | 0  | 0 | 0 | 0 | 0 | 0 | 0 | 0 | 10 | 0.6484 |
|   |      | 7 | 0  | 0 | 0 | 0 | 0 | 0 | 0 | 0 | 0 | 9.3515 |
|   | P.A. |   | 1.6343 | 1.6707 | 2.0084 | 2.3109 | 2.6543 | 2.9765 | 4 | 4.0098 | 5.7614 | 7.3733 |
| 8 | T.A. | 1 | 10 | 7.4864 | 0 | 0 | 0 | 0 | 0 | 0 | 0 | 0 |
|   |      | 2 | 0  | 2.5135 | 10 | 3.7891 | 0 | 0 | 0 | 0 | 0 | 0 |
|   |      | 3 | 0  | 0 | 0 | 6.2108 | 8.9349 | 0 | 0 | 0 | 0 | 0 |
|   |      | 4 | 0  | 0 | 0 | 0 | 1.0650 | 10 | 2.0483 | 0 | 0 | 0 |
|   |      | 5 | 0  | 0 | 0 | 0 | 0 | 0 | 7.9516 | 3.4281 | 0 | 0 |
|   |      | 6 | 0  | 0 | 0 | 0 | 0 | 0 | 0 | 6.5718 | 3.7485 | 0 |
|   |      | 7 | 0  | 0 | 0 | 0 | 0 | 0 | 0 | 0 | 6.2515 | 2.4839 |
|   |      | 8 | 0  | 0 | 0 | 0 | 0 | 0 | 0 | 0 | 0 | 7.5160 |
|   | P.A. |   | 1.4847 | 1.5678 | 1.8273 | 2.1002 | 2.3413 | 2.9323 | 4 | 4.4137 | 5.7667 | 7.9655 |

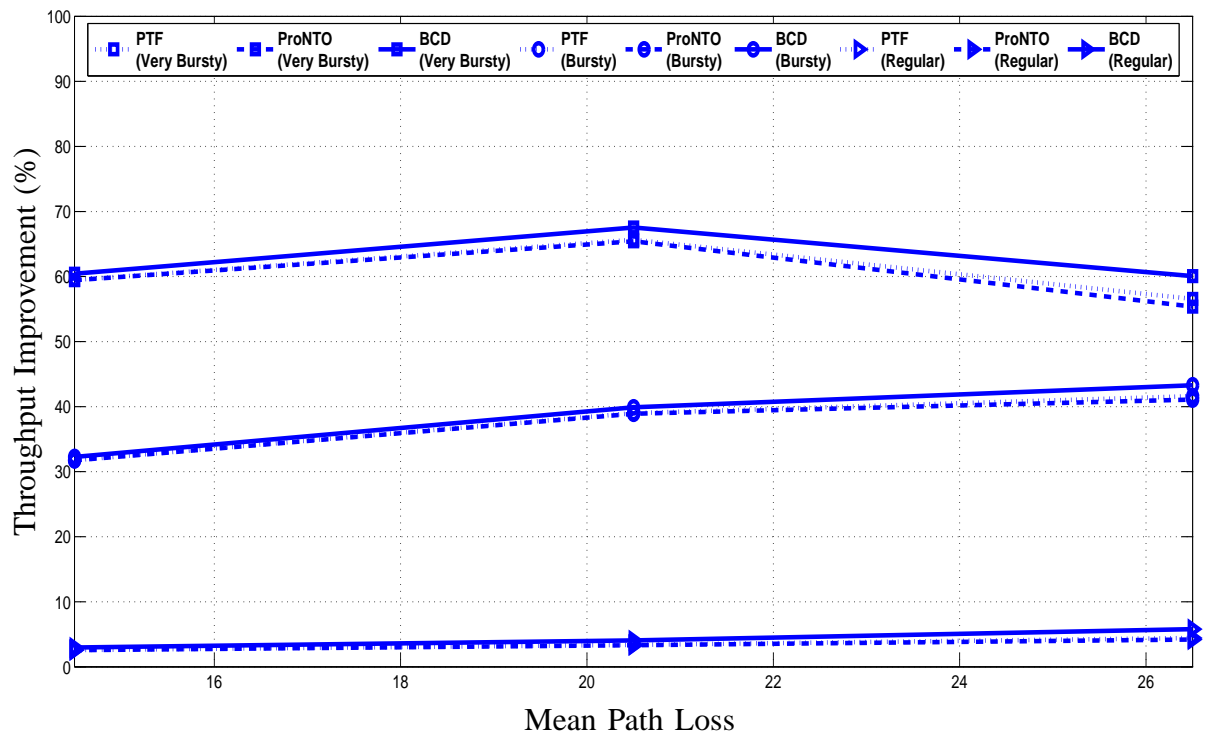

Fig. 3. Throughput Improvement (BCD, PTF, ProNTO) vs. Mean Path Loss for $N = 2$: The effect of mean path loss on throughput improvement for the three energy harvesting cases; *Regular*, *Bursty*, *Very Bursty*.

TABLE IV
FAIRNESS INDEX (SG+TDMA, PTF, PRONTO, BCD) VS. NO. OF USERS: THE FAIRNESS OF PTF AND PRONTO HEURISTICS ARE COMPARED TO THAT OF SG+TDMA'S AND BCD'S, THROUGH $FI$.

| Number of Users | Fairness Index (FI) | | | | | | | | | | | |
|---|---|---|---|---|---|---|---|---|---|---|---|---|
| | Regular | | | | Bursty | | | | Very Bursty | | | |
| | SG+TDMA | PTF | ProNTO | BCD | SG+TDMA | PTF | ProNTO | BCD | SG+TDMA | PTF | ProNTO | BCD |
| 2 | 0.9989 | 0.9949 | 0.9997 | 0.9911 | 1.0000 | 0.9944 | 0.9998 | 0.9915 | 0.9079 | 0.9844 | 0.9997 | 0.9855 |
| 3 | 0.9667 | 0.9813 | 0.9931 | 0.9744 | 0.8079 | 0.9972 | 0.9501 | 0.9672 | 0.6398 | 0.9755 | 0.9633 | 0.9635 |
| 4 | 0.9333 | 0.9484 | 0.9781 | 0.9439 | 0.6520 | 0.9018 | 0.8917 | 0.9288 | 0.8035 | 0.8822 | 0.9642 | 0.9032 |
| 5 | 0.7487 | 0.9650 | 0.8568 | 0.9034 | 0.5554 | 0.8969 | 0.9360 | 0.8921 | 0.5764 | 0.8913 | 0.8308 | 0.8455 |
| 6 | 0.8425 | 0.8291 | 0.9059 | 0.8426 | 0.5594 | 0.7804 | 0.7842 | 0.8147 | 0.3123 | 0.9141 | 0.6706 | 0.7941 |
| 7 | 0.6796 | 0.8567 | 0.7783 | 0.7842 | 0.5399 | 0.8627 | 0.6613 | 0.7115 | 0.1958 | 0.8098 | 0.5695 | 0.7186 |
| 8 | 0.5800 | 0.8172 | 0.6582 | 0.7100 | 0.3554 | 0.7736 | 0.5627 | 0.6355 | 0.2456 | 0.6257 | 0.6915 | 0.6466 |

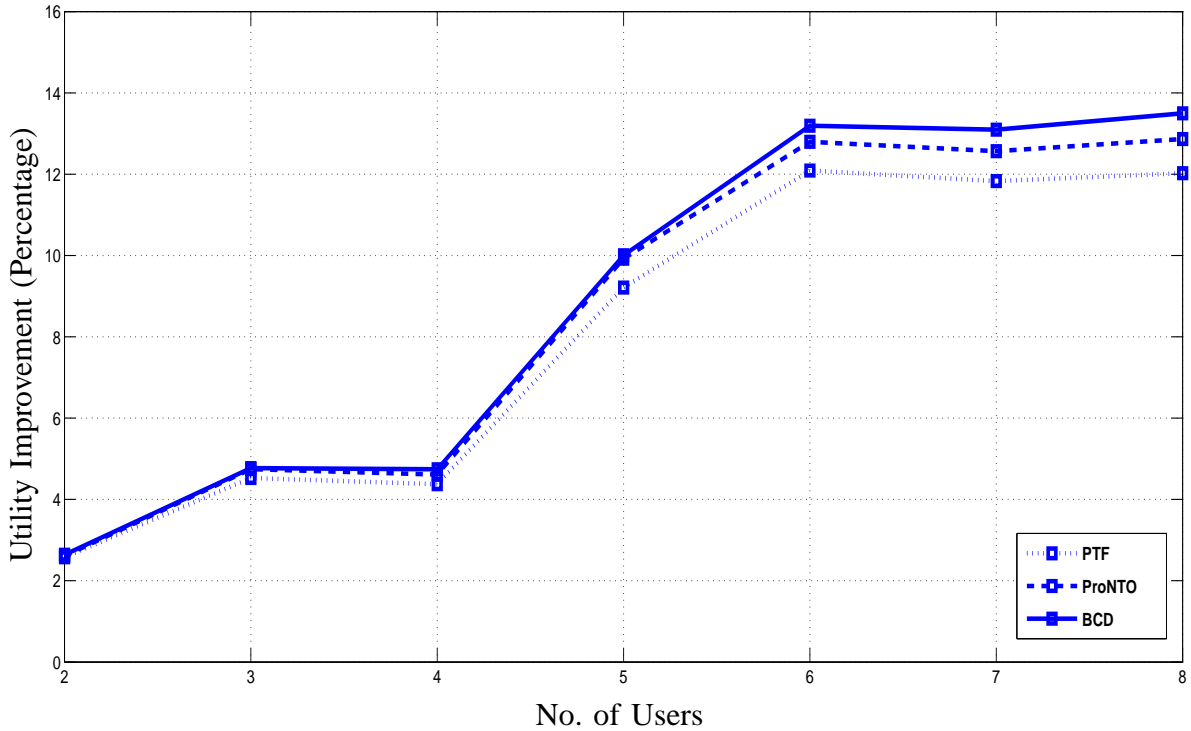

Fig. 4. Average Utility Improvement (PTF, ProNTO, BCD) vs. No. of Users: The average is taken over *Regular*, *Bursty*, *Very Bursty* cases. The average utility improvements of the proposed algorithms over SG+TDMA, for increasing number of users, are compared. Utility improvment increases with increasing number of users.

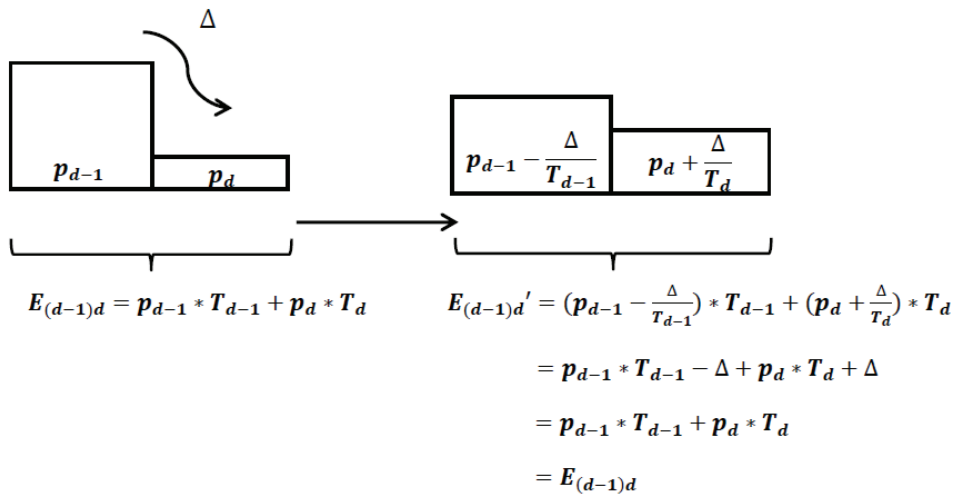

Fig. 5. Maintaining Energy Causality After Energy Deferral